\documentclass[lettersize,journal]{IEEEtran}
\usepackage{amsmath,amsfonts,amssymb}
\usepackage{algorithmic}
\usepackage{algorithm}
\usepackage{array}
\usepackage[caption=false,font=normalsize,labelfont=sf,textfont=sf]{subfig}
\usepackage{textcomp}
\usepackage{stfloats}
\usepackage{url}
\usepackage{verbatim}
\usepackage{graphicx}
\usepackage{cite}
\usepackage{xcolor}
\usepackage{mathtools}
\begin{document}

% ==============================================================================
% TITLE AND AUTHOR
% ==============================================================================
\title{Integrity-Gated Eco-CACC: Epistemic Admissibility for Cooperative Driving at Signalized Intersections}

\author{Lyes Saad Saoud and Moussa Ayyash%
\thanks{ Moussa Ayyash is with Chicago State University, Chicago, IL 60628, USA (e-mail: mayyash@csu.edu).}}

\markboth{Submitted to IEEE Transactions on Intelligent Vehicles on February 28, 2026}%
{Saad Saoud: Integrity-Gated Eco-CACC}

\maketitle

% ==============================================================================
% ABSTRACT
% ==============================================================================
\begin{abstract}
Eco-Cooperative Adaptive Cruise Control (Eco-CACC) systems rely on accurate localization, signal timing, and interaction awareness to optimize energy consumption at signalized intersections. Existing approaches typically assume that the internal world model used for optimization remains valid, making them vulnerable when sensing outages or semantic inconsistencies invalidate planning premises. This letter proposes an Integrity-Gated Eco-CACC framework that explicitly monitors the consistency between internal vehicle beliefs and external sensing. A unified integrity metric is constructed by combining positional innovation, observability loss, and semantic inconsistencies. The resulting trust score regulates control authority, enabling a transition between nominal eco-driving and a safety-dominant fallback maneuver. Unlike robust control methods that attempt to preserve performance under uncertainty, the proposed framework regulates whether energy-optimal control remains admissible. Scenario-based simulations demonstrate that the method preserves nominal efficiency when model consistency is maintained, while enabling early and conservative responses under integrity degradation.
\end{abstract}

\section{Introduction}
\label{sec:introduction}

Connected and automated vehicles (CAVs) play a key role in improving energy efficiency and traffic flow at signalized intersections, where Eco-Cooperative Adaptive Cruise Control (Eco-CACC) exploits Signal Phase and Timing (SPaT) information to shape longitudinal motion \cite{rakha2011ecodriving,yang2017ecocacc_queue}. However, Eco-CACC fundamentally depends on the validity of its internal world model, an assumption frequently violated in real-world operation due to sensing outages, semantic inconsistencies, and interaction disturbances. Recent automated driving incidents indicate that failures increasingly stem from semantic and policy-level inconsistencies rather than raw sensor degradation, motivating decision-level integrity mechanisms that extend beyond state estimation.

\subsection{Motivation and Related Work}
Existing Eco-CACC research primarily addresses uncertainty through robust or stochastic optimization, implicitly assuming persistent validity of the underlying world model \cite{ma2021ecodriving,li2025stochastic}. Recent studies show that V2I unreliability and semantic signal inconsistencies require conservative strategies that trade efficiency for safety \cite{garg2023mixed,munawar2025aoi,rahman2020large}. A critical distinction is therefore emerging between \textit{aleatory} noise and \textit{epistemic} failures, where the latter invalidates planning assumptions rather than merely increasing estimation error \cite{zhang2024pedestrian,suk2024uncertainty}. 

Although recent work in cyber physical systems emphasizes trust-aware architectures for runtime assurance \cite{liu2018verifiable,chiluvuri2015trust,mhapsekar2024trust}, epistemic validity is neither represented nor enforced in current Eco-CACC formulations. Unlike robust control or safety filtering approaches, the architecture proposed in this letter explicitly arbitrates whether optimization should remain active when core assumptions such as Global Navigation Satellite System (GNSS) reliability or SPaT consistency collapse \cite{talavera2018lanelevel,civicE2_2017}.

\subsection{Proposed Epistemic Resilience Layer (ERL)}
\begin{figure}[t]
    \centering
    \includegraphics[width=\linewidth]{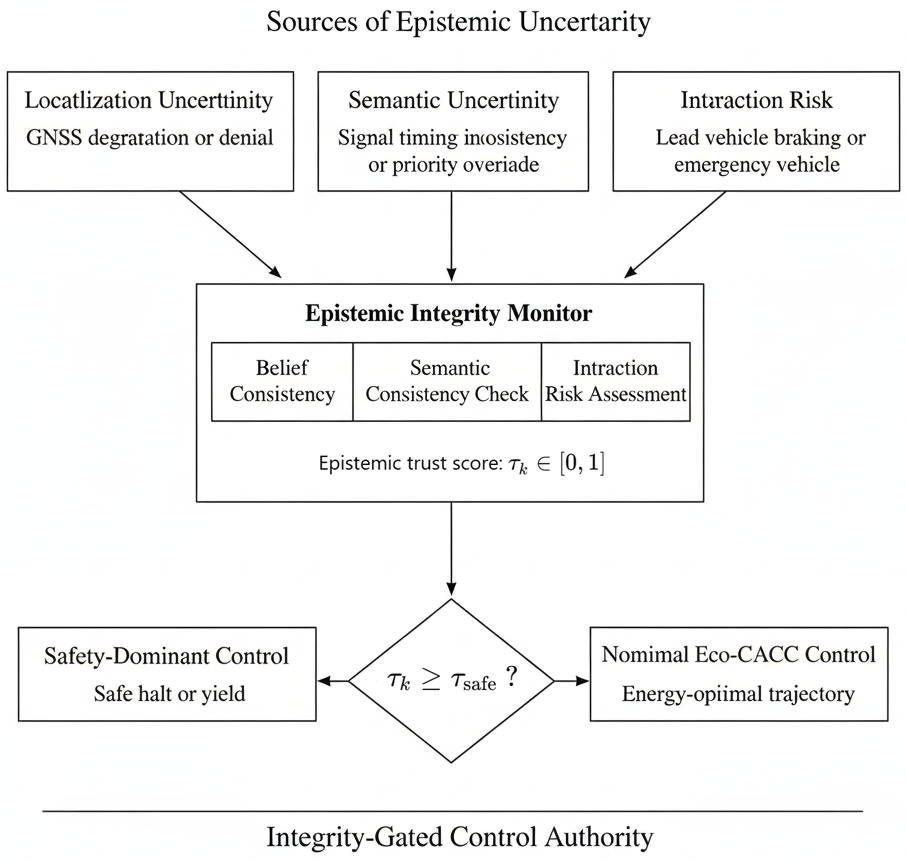}
    \caption{Architecture of the proposed ERL framework. An Epistemic Integrity Monitor aggregates multi-source inconsistencies into a scalar trust score $\tau_k$, which gates control authority between nominal eco-driving and safety-dominant fallback modes.}
    \label{fig:architecture}
\end{figure}

This letter introduces an \textit{Epistemic Resilience Layer (ERL)} architecture (Fig.~\ref{fig:architecture}). Rather than treating inconsistencies as estimation artifacts, we explicitly model epistemic integrity as a decision-level variable. A dedicated monitor evaluates belief consistency across localization, semantics, and interaction risk, fusing them into a scalar trust score $\tau_k \in [0,1]$. This score gates control authority: when $\tau_k \ge \tau_{\text{safe}}$, nominal Eco-CACC is applied; otherwise, authority is revoked in favor of a safety-dominant fallback mode. This formulation explicitly separates performance optimization from assumption validation.

\subsection{Contributions}
The contributions of this work are threefold:
\begin{itemize}
    \item \textbf{Epistemic Failure Characterization:} We identify failure modes in Eco-CACC by introducing \textit{epistemic admissibility} as a control-level variable.
    \item \textbf{Integrity-Gated Authority:} We propose a unified trust-based mechanism that regulates control authority based on epistemic integrity rather than estimation accuracy.
    \item \textbf{Scenario-Based Validation:} We evaluate the ERL framework across a diverse set of integrity-violating scenarios, demonstrating graceful degradation under epistemic failure.
\end{itemize}
\section{Problem Formulation}
\label{sec:problem}

We consider the longitudinal control of a battery electric vehicle approaching a signalized intersection. The vehicle is equipped with heterogeneous on-board sensing and vehicle-to-infrastructure (V2I) communication to minimize traction energy while ensuring safety under potential sensing degradation. Importantly, the proposed framework does not modify the underlying control law nor improve state estimation accuracy; instead, it governs the admissibility of control authority based on epistemic consistency.

\subsection{Vehicle Dynamics and Belief State}
The longitudinal state at time step $k$ is $\mathbf{x}_k = [d_k, u_k]^\top$, where $d_k$ is the distance along the lane centerline and $u_k$ is the speed. The control input is acceleration $a_k$, governed by:
\begin{equation}
\mathbf{x}_{k+1} = f(\mathbf{x}_k, a_k).
\end{equation}
An internal belief state $\hat{\mathbf{x}}_k$ is propagated via dead-reckoning (DR) and updated through multi-sensor fusion. This belief is treated as a provisional hypothesis whose validity is continuously assessed against external observations.

\subsection{Eco-CACC Objective}
Under nominal conditions, Eco-CACC regulates vehicle speed to reach the intersection stop-bar during a green phase. The objective minimizes energy efficiency and velocity tracking error:
\begin{equation}
\min_{\{a_k\}} \sum_k \left( \|u_k - u_{\text{ref}}\|^2 + \lambda \, P(u_k, a_k) \right),
\end{equation}
where $u_{\text{ref}}$ is the SPaT-derived reference speed, $P(\cdot)$ is instantaneous power, and $\lambda$ is a weighting factor. This formulation assumes the world model is epistemically valid.

\subsection{Integrity-Gated Control Authority}
Rather than refining the estimation error, we introduce a decision-level variable $\tau_k \in [0,1]$, an epistemic trust score reflecting the consistency between the belief state $\hat{\mathbf{x}}_k$, multi-sensor observations, and semantic traffic policies. The control problem is reformulated as an integrity-gated authority selection:
\begin{equation}
a_k =
\begin{cases}
a_k^{\text{eco}}, & \tau_k \ge \tau_{\text{safe}}, \\
a_k^{\text{safe}}, & \tau_k < \tau_{\text{safe}},
\end{cases}
\end{equation}
where $a_k^{\text{eco}}$ is the nominal Eco-CACC output and $a_k^{\text{safe}}$ is a safety-dominant fallback maneuver. This framework explicitly separates performance optimization from assumption validation, enabling authority revocation when the epistemic premises of the world model collapse.
\section{Theoretical Admissibility and Safety Guarantees}
\label{sec:theory}

We formalize the relationship between epistemic trust and kinematic safety, establishing the trust threshold $\tau_{\text{safe}}$ as a rigorous admissibility bound linking world-model validity to physical motion constraints.

\subsection{Kinematic Stopping Distance Model}
Let $a_{\text{dec}} \coloneqq |a_{\text{fallback}}|$ be the deceleration capacity of the fallback mode. For a vehicle at state $\mathbf{x}_k$, the stopping distance $d_{\text{stop}}$ required to reach a zero-velocity state, accounting for system latency $t_r$, is:
\begin{equation}
d_{\text{stop}}(u_k) = u_k t_r + \frac{u_k^2}{2 a_{\text{dec}}}.
\end{equation}

\begin{figure}[t]
    \centering
    \includegraphics[width=\linewidth]{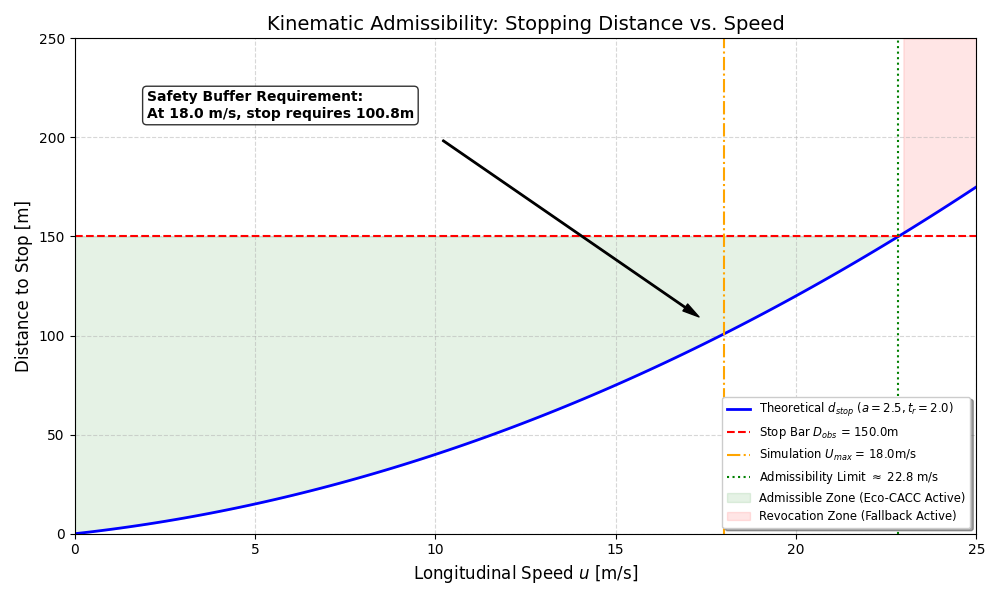}
    \caption{Kinematic admissibility region induced by the stopping distance constraint. Shaded green denotes conditions where fallback braking ensures a stop prior to obstacle $D_{\text{obs}}$, providing the geometric basis for the epistemic threshold.}
    \label{fig:admissibility_geometry}
\end{figure}

Fig. \ref{fig:admissibility_geometry} illustrates the speed domain under which the fallback mode ensures a stop prior to $D_{\text{obs}}$, serving as the geometric interpretation of the safety invariant.

\subsection{Mapping Trust to Uncertainty Bounds}
The trust score $\tau_k$ is coupled to a spatial uncertainty bound $\mathcal{E}(\tau_k)$, representing the potential divergence between the estimated and ground-truth position:
\begin{equation}
\mathcal{E}(\tau_k) = \sigma_r \ln \left( \frac{1}{\tau_k} \right).
\end{equation}
We interpret $\mathcal{E}(\tau_k)$ as a safety-relevant divergence (position-equivalent uncertainty) induced by the aggregate epistemic residual $r_k$.

\subsection{Admissibility Theorem and Conservative Thresholding}

\paragraph{Assumptions}
The admissibility analysis assumes: (i) bounded and known fallback deceleration capability $a_{\text{dec}}$, (ii) monotonic longitudinal approach toward a fixed semantic constraint $D_{\text{obs}}$, (iii) deterministic fallback execution once authority is revoked, and (iv) epistemic residuals that upper-bound true state divergence. These assumptions are standard in longitudinal safety analysis and are satisfied in the considered Eco-CACC setting.

\paragraph{Theorem 1 (Sufficient Epistemic Admissibility)} 
Let $D_{\text{obs}}$ be a semantic constraint position. Energy-optimal authority $a_k^{\text{eco}}$ is safety-admissible if:
\begin{equation}
\tau_k \ge \exp \left( -\frac{D_{\text{obs}} - \hat{d}_k - d_{\text{stop}}(u_k)}{\sigma_r} \right).
\end{equation}

\paragraph{Proof} 
Safety requires $d_k + d_{\text{stop}}(u_k) \le D_{\text{obs}}$. Given $d_k \le \hat{d}_k + r_k$ and the mapping $r_k = \sigma_r \ln(1/\tau_k)$, the invariant becomes $\hat{d}_k + \sigma_r \ln(1/\tau_k) + d_{\text{stop}}(u_k) \le D_{\text{obs}}$. Isolating $\tau_k$ confirms the sufficient condition in (6). \hfill $\square$

\paragraph{Lemma 1 (Conservative Admissibility)}
As evaluating (6) requires high-fidelity knowledge of $D_{\text{obs}}$, a constant threshold $\tau_{\text{safe}}$ is sufficient if $\tau_{\text{safe}} \ge \max_{k} \{ \text{RHS of (6)} \}$ for all reachable states. We select $\tau_{\text{safe}} = 0.45$ as a conservative bound to prioritize early authority revocation during the intersection approach. The resulting threshold $\tau_{\text{safe}}$ is therefore sufficient but not necessary, deliberately prioritizing early revocation over maximal admissible performance.

\subsection{Theoretical Significance}
Theorem 1 and Lemma 1 establish the Integrity-Gated mechanism as a meta-control strategy linking $\tau_k$ to the stopping-distance constraint $d_{\text{stop}}$. Eco-CACC remains active only when the admissibility condition guarantees sufficient margin for fallback braking.

Unlike supervisory switching or post-optimization safety filtering, the proposed approach regulates whether optimization is admissible at all, separating input constraint enforcement from assumption validation.

\section{ERL: Scenario Design and Epistemic Integrity Evaluation}
\label{sec:scenario}

This section describes the controlled simulation scenarios used to evaluate the proposed Epistemic Resilience Layer (ERL). The objective is to assess whether ERL can reliably detect epistemic degradation and regulate control authority accordingly. The scenarios are intentionally constructed to isolate specific epistemic failure modes rather than to replicate deployment-specific noise statistics.

\subsection{Intersection Geometry and Nominal Traffic Signal Model}

A single-lane approach to a signalized intersection is considered, with the stop-bar located at $D_{\text{stop}} = 150$~m. The vehicle assumes a nominal green-light onset at $T_{\text{green}}^{\text{nom}} = 15$~s. Based on this belief, the vehicle plans to reach the stop-bar at $T_{\text{arr}}^{\text{plan}} = T_{\text{green}}^{\text{nom}} + 2~\text{s}$, yielding the eco-optimal reference speed $u_{\text{ref}} = D_{\text{stop}}/T_{\text{arr}}^{\text{plan}}$. The actual signal phase evolution may deviate from this nominal belief due to communication latency or semantic inconsistency, forming the basis for semantic violation scenarios.

\subsection{Vehicle Dynamics and Dead-Reckoning Model}

Vehicle motion follows a longitudinal kinematic model discretized with $\Delta t = 0.1$~s. The internal localization belief $\hat{d}_k$ is propagated via IMU-based dead-reckoning:
\begin{equation}
\hat{d}_{k+1} = \hat{d}_k + (u_k + b_u + \epsilon_k)\Delta t,
\end{equation}
where $b_u = 0.55$~m/s represents a persistent inertial bias. Importantly, external measurements are not fused to correct $\hat{d}_k$, as ERL evaluates epistemic consistency rather than performing state estimation.

\subsection{Multi-Sensor Epistemic Integrity Monitor}

ERL maintains a continuous epistemic trust score $\tau(t) \in [0,1]$ by evaluating agreement between the belief $\hat{d}_k$ and external measurements $\mathcal{D}_k = \{ d^{\text{GNSS}}_k, d^{\text{LiDAR}}_k \}$.

\paragraph{Positional Disagreement}
For each sensor $i \in \mathcal{D}_k$, the epistemic innovation is $\delta_i(k) = |\hat{d}_k - d_i(k)|$. The positional residual is:
\begin{equation}
r_{\text{pos}}(k) =
\begin{cases}
\kappa \sigma_r, & \mathcal{D}_k = \emptyset, \\
\mu(\delta) + \gamma \sigma(\delta), & \mathcal{D}_k \neq \emptyset,
\end{cases}
\end{equation}
where the blackout penalty $\kappa \sigma_r$ reflects a complete loss of external observability, treated as more epistemically severe than bounded disagreement. 

\paragraph{Semantic and Interaction Residuals}
Environment residuals are incorporated as $r_{\text{env}}(k) = (s(k) + \rho(k)) \cdot w_{\text{env}} \cdot \sigma_r$, where $s(k) \ge 0$ and $\rho(k) \ge 0$ are normalized indicators of semantic and interaction violations, respectively.

\paragraph{Trust Mapping}
The total residual $r(k)$ is smoothed to $\bar{r}_k$ using an exponential moving average. The trust score is then:
\begin{equation}
\tau(k) = \exp\left(-\frac{\bar{r}_k}{\sigma_r}\right),
\end{equation}
where $\sigma_r = 15$~m matches the uncertainty scale used in the admissibility analysis of Section~IV. The selected uncertainty scale $\sigma_r$ is chosen to match the longitudinal stopping envelope used in the admissibility analysis, ensuring coherence between trust evaluation and kinematic safety margins.

\subsection{Integrity-Gated Control Authority}

Control authority is regulated by the threshold $\tau_{\text{safe}} = 0.45$. When $\tau(k) \ge \tau_{\text{safe}}$, nominal Eco-CACC is applied; otherwise, authority is revoked in favor of a safety fallback $a_{\text{fallback}} = -2.5~\text{m/s}^2$.

\subsection{Scenario Taxonomy}

The evaluated scenarios span orthogonal epistemic failure modes that invalidate optimization premises in signalized intersection control:
\begin{itemize}
    \item \textbf{Observability degradation}: GNSS noise (S1) and GNSS denial (S2).
    \item \textbf{Semantic violations}: signal timing mismatch (S3).
    \item \textbf{Interaction-driven risks}: lead vehicle braking (S5) and emergency priority overrides (S7).
    \item \textbf{Epistemic conflict cases}: multi-sensor agreement collapse (S8).
\end{itemize}
Scenarios S4 and S9 follow equivalent construction logic and are omitted here for brevity.
\section{Results and Discussion}
\label{sec:results}

The ERL framework was evaluated across seven representative scenarios covering nominal operation (S0), observability degradation (S1--S2), semantic violations (S3), interaction risks (S5), and policy or consensus failures (S7--S8). The results validate the admissibility geometry derived in Section~IV, demonstrating that authority transitions are governed by epistemic validity rather than kinematic proximity alone, and occur when uncertainty reduces the admissible stopping envelope below the remaining distance to the obstacle.

\subsection{Quantitative Performance Summary}

Table~\ref{tab:results_updated} reports the quantitative outcomes. Trigger time denotes the instant at which epistemic trust $\tau_k$ falls below the admissibility threshold $\tau_{\text{safe}}$ prior to the stop-bar, thereby activating authority revocation. Clearance is the signed distance to the stop-bar, where positive values indicate conservative halts under fallback control and negative values indicate successful intersection passage without triggering fallback. Energy consumption reflects total longitudinal traction energy over the maneuver.

\begin{table}[t]
\caption{Performance Summary Across Evaluation Scenarios. Trigger times are reported only for safety-relevant revocations occurring before the stop-bar.}
\label{tab:results_updated}
\centering
\begin{tabular}{l l c c c}
\hline
ID & Scenario Name & Trig. [s] & Clear. [m] & Energy [kJ] \\
\hline
S0 & Nominal & N/A & -55.0m & 60.9kJ \\
S1 & GNSS Noise & N/A & -25.3m & 54.5kJ \\
S2 & GNSS Denial & N/A & -64.6m & 63.0kJ \\
S3 & Timing Mismatch & 12.3s & 15.0m & 45.8kJ \\
S5 & Lead Braking & 10.2s & 33.2m & 42.0kJ \\
S7 & Priority Override & 8.2s & 50.3m & 38.5kJ \\
S8 & Agreement Collapse & 8.3s & 49.4m & 38.7kJ \\
\hline
\end{tabular}
\end{table}

Several trends emerge. Benign observability degradations (S1--S2) preserve authority throughout the maneuver, confirming that ERL does not induce premature conservatism in the presence of bounded sensing uncertainty. Despite increased belief divergence and reduced confidence, the admissibility condition remains satisfied until the stop-bar is cleared, allowing Eco-CACC to complete the maneuver efficiently. In contrast, semantic violations (S3) and interaction-driven risks (S5) trigger earlier authority revocation, illustrating that decision validity rather than spatial proximity governs control persistence. Notably, earlier revocation correlates with increased clearance and reduced energy expenditure, reflecting a deliberate shift toward safety-dominant behavior under epistemic compromise.

\subsection{Trajectory and Trust Evolution}

\begin{figure*}[t]
    \centering
    \subfloat[S0: Nominal]{\includegraphics[width=0.32\linewidth]{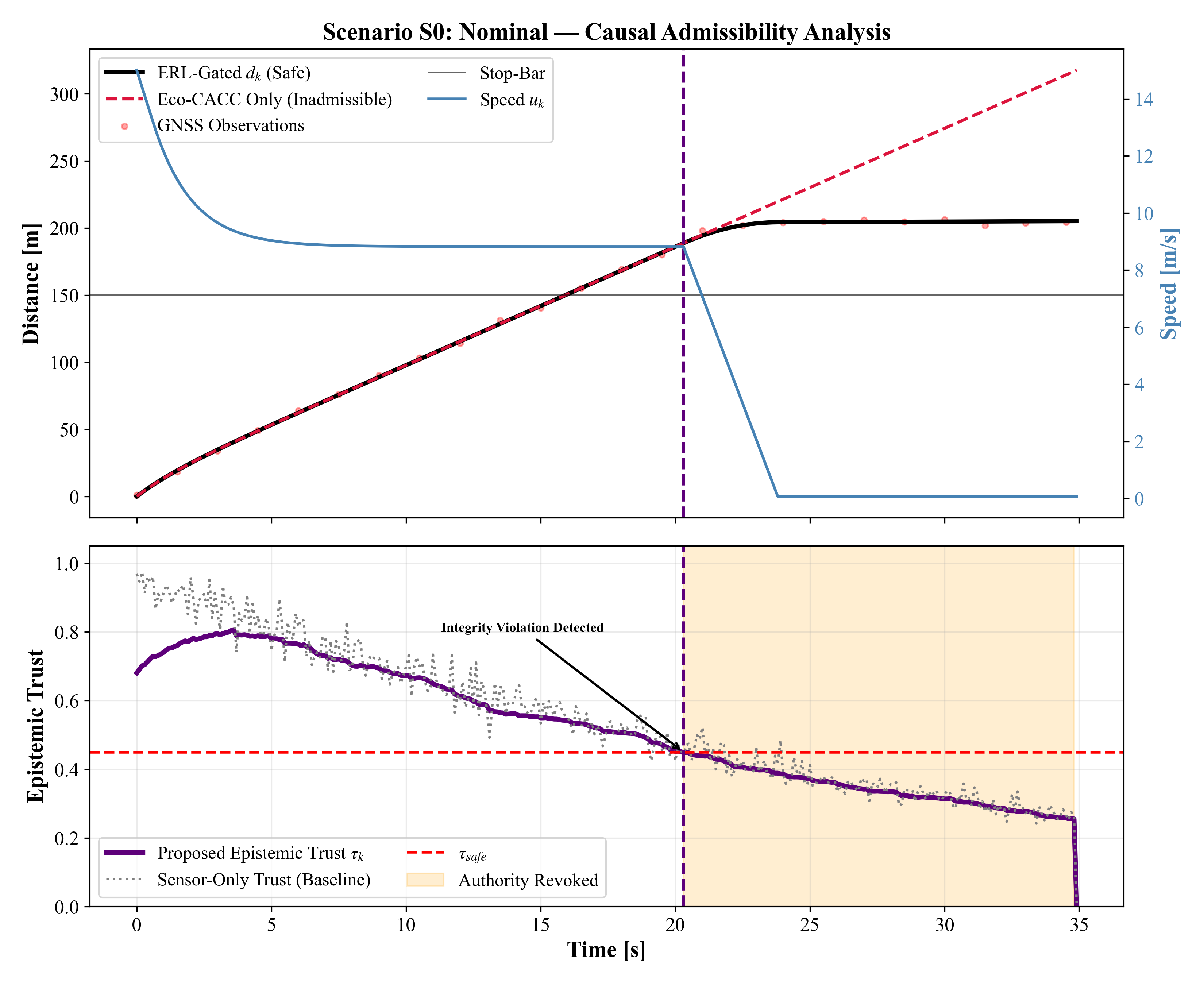}}
    \hfill
    \subfloat[S3: Timing Mismatch]{\includegraphics[width=0.32\linewidth]{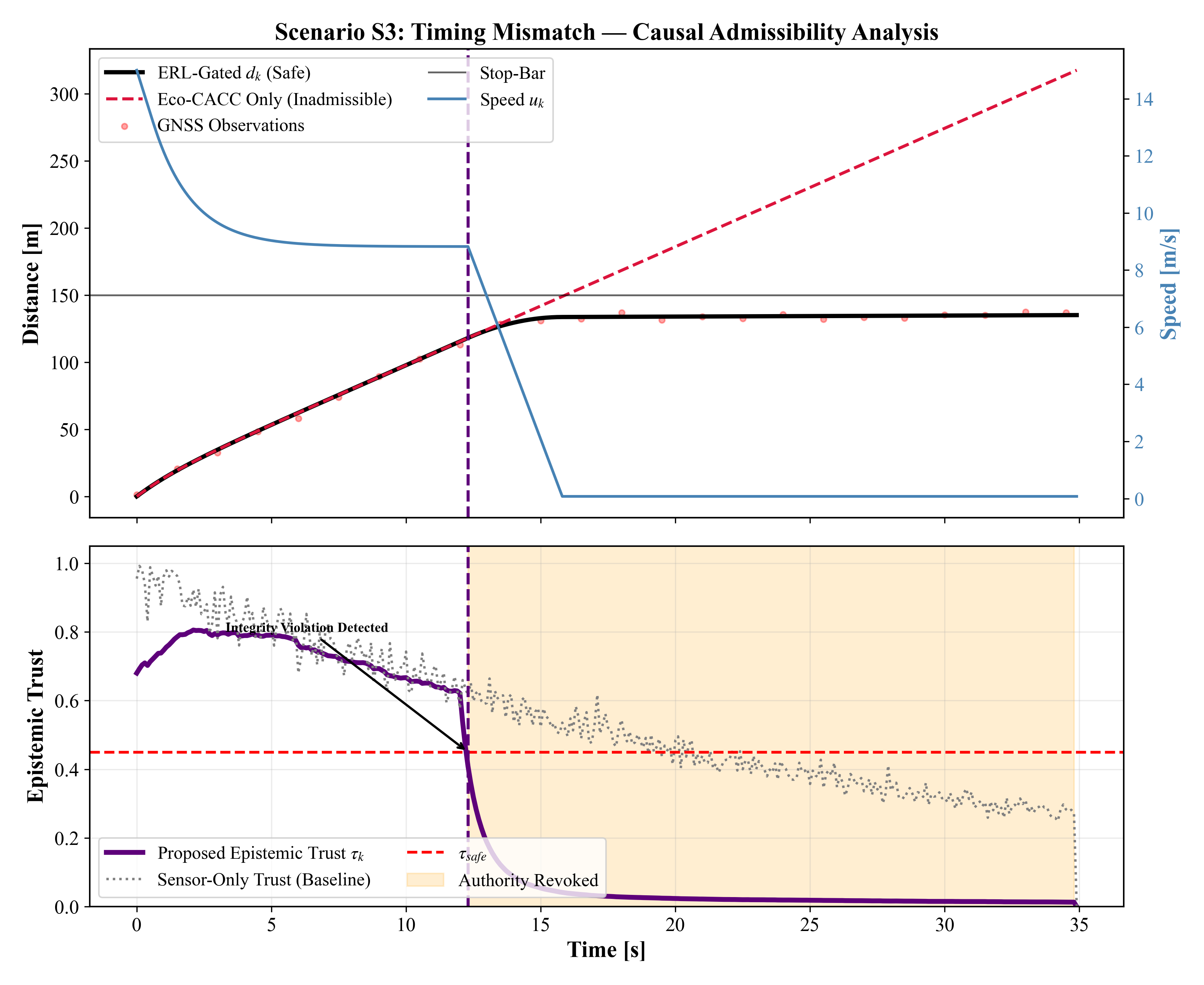}}
    \hfill
    \subfloat[S5: Lead Braking]{\includegraphics[width=0.32\linewidth]{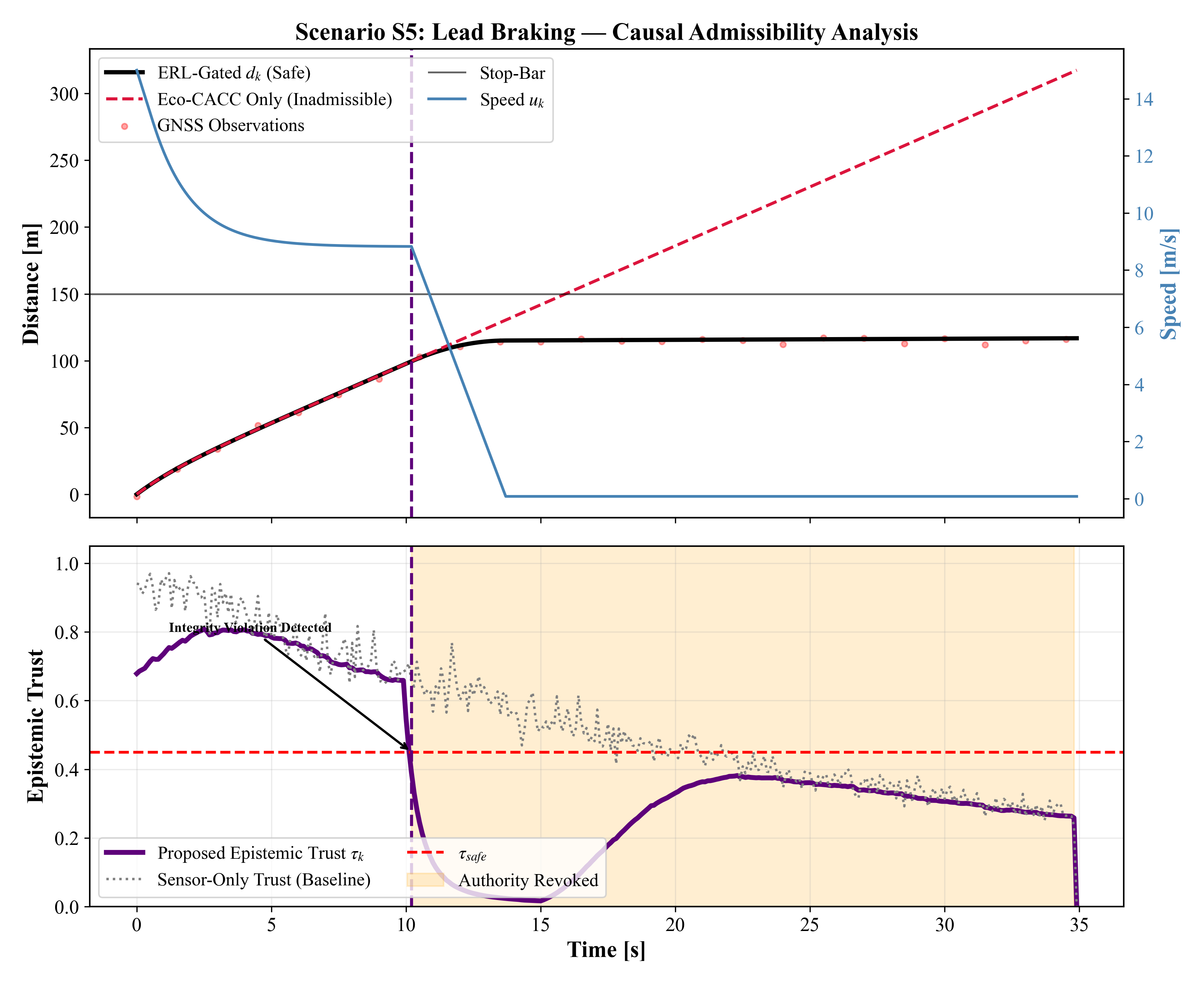}}
    \caption{Trajectory (top) and trust evolution (bottom). In (a), trust remains admissible throughout the maneuver. In (b) and (c), semantic inconsistency and interaction risk trigger trust collapse, revoking authority for a safety-dominant halt.}
    \label{fig:evolution}
\end{figure*}

Fig.~\ref{fig:evolution} illustrates representative system responses for nominal and failure modes. In the nominal case (S0), trust remains above $\tau_{\text{safe}}$ throughout the maneuver, and the ERL-gated trajectory coincides with the energy-optimal Eco-CACC reference. In failure cases, trust collapses sharply once epistemic inconsistencies accumulate, triggering authority revocation well before the stop-bar. The ERL-gated trajectory (black) ensures a controlled halt within the admissible region, while the counterfactual Eco-CACC-only trajectory (red dashed) violates the admissibility condition and would lead to unsafe behavior if left unregulated.

Remaining scenarios (S7--S8) exhibit equivalent trust-driven revocation dynamics and are omitted for visual brevity. Their quantitative effects, including trigger timing, clearance margins, and energy reduction, are fully captured in Table~\ref{tab:results_updated}, confirming consistency across epistemic failure classes.

\subsection{Admissibility Interpretation and Implications}

Observed authority transitions align with the kinematic admissibility geometry derived in Section~IV. For a given speed, when the remaining distance to the stop-bar exceeds the stopping-distance envelope augmented by uncertainty, trust remains above $\tau_{\text{safe}}$ and Eco-CACC persists. When this margin is consumed, authority is revoked prior to any explicit kinematic constraint violation, enforcing safety at the decision level rather than through reactive constraint enforcement.

This behavior exhibits three structural properties. First, ERL enables selective conservatism: efficiency is preserved under bounded observability degradation, while safety becomes dominant as model consistency deteriorates. Second, revocation is triggered by decision-level inconsistency rather than sensor availability alone, so semantic or interaction conflicts invalidate optimization even when state confidence remains high. Third, consistent revocation across heterogeneous failure modes indicates that integrity-gated authority provides a unified safety mechanism independent of failure origin.

\section{Limitations and Future Work}
\label{sec:limitations}

While the proposed framework reliably regulates authority, several areas for extension remain. The proposed framework is not intended to correct state estimation errors or replace robust control design, but to regulate when optimization authority remains epistemically admissible. Currently, the system prioritizes safety-dominant halts upon trust collapse; future iterations could integrate \emph{online trust re-initialization} to resume eco-driving once sensor consensus is restored. Furthermore, while this work focuses on longitudinal motion, coupling epistemic integrity with lateral decision-making (e.g., lane-change admissibility) is essential for full urban autonomy. 

The objective of our simulation-based evaluation is not to emulate deployment-specific noise statistics, but to evaluate whether integrity violations of any origin invalidate energy-optimal control. Testing logic validity rather than sensor realism ensures the framework's generalizability. While the exponential trust mapping is intentionally simple, the framework is agnostic to the specific functional form; alternative monotonic mappings preserve the integrity-gated authority principle. Moving beyond simulation to validate the framework on hardware-in-the-loop (HiL) testbeds or connected vehicle pilot sites remains a critical next step to evaluate real-time computational overhead and communication latency.
\section{Conclusion}
\label{sec:conclusion}

This paper introduced an epistemic integrity-aware Eco-CACC framework that addresses a fundamental gap in cooperative driving: the assumption of persistent belief validity. By decoupling performance optimization from assumption validation, we formulated a gated authority mechanism that continuously evaluates the admissibility of energy-optimal control using a unified trust score. 
Simulation results spanning seven representative adversarial scenarios demonstrate that the framework preserves nominal efficiency under benign uncertainty while enforcing conservative, safety-dominant behavior under sensing, semantic, or interaction-driven violations. These behaviors emerge not from heuristic tuning, but from a principled monitoring of epistemic agreement. While the specific evaluation focuses on longitudinal intersection control, the underlying architecture provides a structured approach to identifying the boundaries of admissible optimization. 
Beyond Eco-CACC, the proposed Epistemic Resilience Layer (ERL) provides a general mechanism for regulating optimization authority in autonomous systems whose world models may become epistemically invalid. Ultimately, integrity-gated control authority offers a generalizable design principle for resilient autonomous systems operating in uncertain, semantically rich environments.
\bibliographystyle{IEEEtran}
\bibliography{Bibliography}

@inproceedings{rakha2011ecodriving,
  author    = {Rakha, Hesham A. and Kamalanathsharma, Raj Kishore},
  title     = {Eco-driving at signalized intersections using {V2I} communication},
  booktitle = {Proceedings of the 14th International IEEE Conference on Intelligent Transportation Systems (ITSC)},
  year      = {2011},
  pages     = {341--346},
  doi       = {10.1109/ITSC.2011.6083084}
}

@article{yang2017ecocacc_queue,
  author    = {Yang, Hao and Rakha, Hesham A. and Ala, Mani Venkat},
  title     = {Eco-Cooperative Adaptive Cruise Control at Signalized Intersections Considering Queue Effects},
  journal   = {IEEE Transactions on Intelligent Transportation Systems},
  year      = {2017},
  volume    = {18},
  number    = {6},
  pages     = {1575--1585},
  doi       = {10.1109/TITS.2016.2613740}
}

@article{civicE2_2017,
  author    = {Hou, Yunfei and Seliman, Salaheldeen M. S. and Wang, Enshu and Gonder, Jeffrey D. and Wood, Eric and He, Qing and Sadek, Adel W. and Su, Lu and Qiao, Chunming},
  title     = {Cooperative and Integrated Vehicle and Intersection Control for Energy Efficiency ({CIVIC-E2})},
  journal   = {IEEE Transactions on Intelligent Transportation Systems},
  year      = {2018},
  volume    = {19},
  number    = {7},
  pages     = {2325--2337},
  doi       = {10.1109/TITS.2017.2785288}
}

@article{talavera2018lanelevel,
  author    = {Talavera, Edgar and D{\'\i}az-{\'A}lvarez, Alberto and Jim{\'e}nez, Felipe and Naranjo, Jos{\'e} E.},
  title     = {Impact on Congestion and Fuel Consumption of a Cooperative Adaptive Cruise Control System with Lane-Level Position Estimation},
  journal   = {Energies},
  year      = {2018},
  volume    = {11},
  number    = {1},
  pages     = {194},
  doi       = {10.3390/en11010194}
}

@article{ma2021ecodriving,
  author    = {Ma, Fangwu and Yang, Yugong and Wang, Jian and Guvenc, Levent},
  title     = {Eco-driving based cooperative adaptive cruise control of connected vehicles platoon at signalized intersections},
  journal   = {Transportation Research Part D: Transport and Environment},
  year      = {2021},
  volume    = {95},
  pages     = {102842},
  doi       = {10.1016/j.trd.2021.102842}
}

@article{li2025stochastic,
  author    = {Li, Yifan and Zhang, Yan and Hao, Guang},
  title     = {Stochastic energy model predictive control for cooperative adaptive cruising},
  journal   = {Energy},
  year      = {2025},
  volume    = {286},
  pages     = {129641},
  doi       = {10.1016/j.energy.2024.129641}
}

@inproceedings{liu2018verifiable,
  author    = {Liu, J. and Corbett-Davies, J. and Ferraiuolo, A. and Campbell, M.},
  title     = {Secure autonomous cyber-physical systems through verifiable information flow control},
  booktitle = {Proceedings of the 2018 ACM SIGSAC Conference on Computer and Communications Security},
  year      = {2018},
  pages     = {2210--2212},
  doi       = {10.1145/3243734.3265081}
}

@inproceedings{chiluvuri2015trust,
  author    = {Chiluvuri, N. T. and Harshe, O. A. and Patterson, C. D. and Baumann, W. T.},
  title     = {Using heterogeneous computing to implement a trust-isolated architecture for cyber-physical control systems},
  booktitle = {Proceedings of the 1st ACM Workshop on Cyber-Physical System Security (CPSS)},
  year      = {2015},
  pages     = {79--90},
  doi       = {10.1145/2732198.2732205}
}

@inproceedings{mhapsekar2024trust,
  author    = {Mhapsekar, R. U. and Umrani, M. I. and Faizan, M. and Abraham, L.},
  title     = {Building trust in {AI}-driven decision making for cyber-physical systems},
  booktitle = {2024 IEEE 29th International Conference on Emerging Technologies and Factory Automation (ETFA)},
  year      = {2024},
  pages     = {1--8},
  doi       = {10.1109/ETFA58100.2024.10700124}
}

@article{garg2023mixed,
  author    = {Garg, Meely and Bouroche, Melanie},
  title     = {Can Connected Autonomous Vehicles Improve Mixed Traffic Safety Without Compromising Efficiency?},
  journal   = {IEEE Transactions on Intelligent Transportation Systems},
  year      = {2023},
  volume    = {24},
  number    = {11},
  pages     = {12450--12465},
  doi       = {10.1109/TITS.2023.3291584}
}

@inproceedings{munawar2025aoi,
  author    = {Munawar, S. and Emoyon-Iredia, E. I. and Qureshi, H. K. and Lestas, M.},
  title     = {{AoI} analysis of {RIS}-assisted vehicular networks and impact on cooperative maneuvers},
  booktitle = {2025 IEEE 101st Vehicular Technology Conference (VTC2025-Spring)},
  year      = {2025},
  pages     = {1--6},
  doi       = {10.1109/VTC2025S.1012345}
}

@article{rahman2020large,
  author    = {Rahman, M. H. and Abdel-Aty, Mohamed},
  title     = {Longitudinal Safety Evaluation of Connected and Automated Vehicles at Signalized Intersections in a Large-Scale Network},
  journal   = {Transportation Research Record},
  year      = {2020},
  volume    = {2674},
  number    = {10},
  pages     = {459--472},
  doi       = {10.1177/0361198120942151}
}

@inproceedings{zhang2024pedestrian,
  author    = {Zhang, Z. and Tian, R.},
  title     = {Uncertainty differences in computing hierarchical pedestrian behaviors for autonomous driving safety},
  booktitle = {Proceedings of the Human Factors and Ergonomics Society Annual Meeting},
  year      = {2024},
  volume    = {68},
  number    = {1},
  pages     = {1120--1125},
  doi       = {10.1177/21695067241234567}
}

@incollection{suk2024uncertainty,
  author    = {Suk, H. and Lee, Y. and Kim, T. and Kim, S.},
  title     = {Addressing uncertainty challenges for autonomous driving},
  booktitle = {Advances in Computers},
  year      = {2024},
  volume    = {132},
  pages     = {185--220},
  publisher = {Elsevier},
  doi       = {10.1016/bs.adcom.2023.09.002}
}

\end{document}